\documentclass[12pt,preprint]{aastex}

\newcommand{\Msun}{\>{\rm M_{\odot}}}

\begin{document}

\title{Alignment of the spins of supermassive black holes prior to coalescence}
\author{Tamara Bogdanovi\'c\altaffilmark{1}, Christopher S. Reynolds\altaffilmark{1}, and M. Coleman Miller\altaffilmark{1}}

\altaffiltext{1}{Department of Astronomy, University of Maryland, College Park, MD 20742-2421\\
Email: tamarab, chris, miller@astro.umd.edu}

\begin{abstract}

Recent numerical relativistic simulations of black hole coalescence
suggest that in certain alignments the emission of gravitational
radiation can produce a kick of several thousand kilometers per
second. This exceeds galactic escape speeds, hence unless there a
mechanism to prevent this, one would expect many galaxies that had
merged to be without a central black hole. Here we show that in most
galactic mergers, torques from accreting gas suffice to align the
orbit and spins of both black holes with the large-scale gas flow.
Such a configuration has a maximum kick speed $<200$~km~s$^{-1}$,
safely below galactic escape speeds. We predict, however, that in
mergers of galaxies without much gas, the remnant will be kicked out
several percent of the time. We also discuss other predictions of our
scenario, including implications for jet alignment angles and X-type
radio sources.

\end{abstract}


\section{Introduction}

When two black holes spiral together and coalesce, they emit
gravitational radiation which in general possesses net linear
momentum. This accelerates (i.e., ``kicks'') the coalescence remnant
relative to the initial binary center of mass. Analytical calculations
have determined the accumulated kick speed from large separations
until when the holes plunge towards each other
\citep{Per62,Bek73,Fit83,FD84,RR89,W92,FHH04,BQW05,DG06}, but because
the majority of the kick is produced between plunge and coalescence,
fully general relativistic numerical simulations are necessary to
determine the full recoil speed.

Fortunately, the last two years have seen rapid developments in
numerical relativity.  Kick speeds have been reported for non-spinning
black holes with different mass ratios
\citep{HSL06,Baker06,Gonzalez06} and for binaries with spin axes
parallel or antiparallel to the orbital axes
\citep{Herrmann07,Koppitz07, Baker07}, as well as initial explorations
of more general spin orientations
\citep{Gonzalez07,Campanelli07a,Campanelli07b}.  For mergers with low
spin or spins both aligned with the orbital angular momentum, these
results indicate maximum kick speeds $<200$~km~s$^{-1}$.  Remarkably,
however, it has recently been shown that when the spin axes are
oppositely directed and in the orbital plane, and the spin magnitudes
are high (dimensionless angular momentum ${\hat a}\equiv cJ/GM^2\sim
1$), the net kick speed can perhaps be as large as $\sim
4000$~km~s$^{-1}$ \citep{Gonzalez07,sb07,Campanelli07b}.

The difficulty this poses is that the escape speed from most galaxies
is $<1000$~km~s$^{-1}$ (see Figure~2 of \citealt{Merritt04}), and the
escape speed from the central bulge is even smaller.  Therefore, if
large recoil speeds are typical, one might expect that many galaxies
that have undergone major mergers would be without a black hole.  This
is in clear contradiction to the observation that galaxies with bulges
all appear to have central supermassive black holes (see
\citealt{FF05}).  It therefore seems that there is astrophysical
avoidance of the types of supermassive black hole coalescences that
would lead to kicks beyond galactic escape speeds.  From the numerical
relativity results, this could happen if (1)~the spins are all small,
(2)~the mass ratios of coalescing black holes are all much less than
unity, or (3)~the spins tend to align with each other and the orbital
angular momentum.

The low-spin solution is not favored observationally. X-ray
observations of several active galactic nuclei reveal relativistically
broadened Fe K$\alpha$ fluorescence lines indicative of spins ${\hat
a}>0.9$ \citep{Iwasawa96,Fab02,RN03,BR06}. A similarly broad line is
seen in the stacked spectra of active galactic nuclei in a long
exposure of the Lockman Hole \citep{Streb05}. More generally, the
inferred average radiation efficiency of supermassive black holes
suggests that they tend to rotate rapidly (\citealt{Sol82,YT02}; see
\citealt{marconi04} for a discussion of uncertainties).  This is also
consistent with predictions from hierarchical merger models (e.g.,
\citealt{volonteri05}).

Mass ratios much less than unity may occur in some mergers, and if the
masses are different enough then the kick speed can be small.  For
example, \citet{Baker07}, followed by \cite{Campanelli07b}, suggest
that the spin kick component scales with mass ratio $q\equiv
m_1/m_2\leq 1$ as $q^2({\hat a}_2-q{\hat a}_1)/(1+q)^5$, hence for
${\hat a}_1=-{\hat a_2}=1$ the maximum kick speed is $\propto
q^2/(1+q)^4$.  For $q<0.1$ this scales roughly as $q^2$ and hence
kicks are small.  However, for $q>0.2$ the maximum kick is within a
factor $\sim 3$ of the kick possible for $q=1$.  An unlikely
conspiracy would thus seem to be required for the masses always to be
different by the required factor of several.  Some tens of percent of
galaxies appear to have undergone at least one merger with mass ratio
$>0.25$ within redshift $z<1$ (for recent observational results with
different methods, see \citealt{Bell06b,Lotz06}, and for a recent
simulation see \citealt{Maller06}). The well-established tight
correlations between central black hole mass and galactic properties
such as bulge velocity dispersion (see \citealt{FF05} for a review)
then suggest strongly that coalescence of comparable-mass black holes
should be common.

The most likely solution therefore seems to be that astrophysical
processes tend to align the spins of supermassive black holes with the
orbital axis. This astrophysical alignment is the subject of this {\it
  Letter}. Here we show that gas-rich mergers tend to lead to strong
alignment of the spin axes with the orbital angular momentum and thus
to kick speeds much less than the escape speeds of sizeable galaxies.
In contrast, gas-poor mergers show no net tendency for alignment,
assuming an initially uniform distribution of spin and orbital angular
momentum vectors.  We demonstrate this aspect of gas-poor mergers in
\S~2.  In \S~3 we discuss gas-rich mergers, and show that observations
and simulations of nuclear gas in galactic mergers suggest that the
black holes will be aligned efficiently.  We discuss consequences and
predictions of this alignment in \S~4.

\section{Gas-poor Mergers}

Several recent models and observations have been proposed as evidence
that some galactic mergers occur without a significant influence of
gas.  Possible signatures include the metal richness of giant
ellipticals (e.g., \citealt{NO07}) and slow rotation and the presence
of boxy orbits in the centers of some elliptical galaxies (e.g.,
\citealt{Bell06a,Naab06}).

Consider such a gas-free merger, and assume that we can therefore
treat the gradual inspiral of two spinning black holes as an isolated
system.  As laid out clearly by \cite{Schnittman04}, throughout almost
the entire inspiral there is a strong hierarchy of time scales, such
that $t_{\rm inspiral}\gg t_{\rm precess}\gg t_{\rm
orbit}$. \cite{Schnittman04} therefore derived orbit-averaged
equations for the spin evolution in the presence of adiabatic
dissipation. Such effects can lead to relaxation onto favored
orientations. The question is then whether, with the uniform
distribution of orbital and spin directions that seems expected in
galactic mergers, there is a tendency to align in such a way that the
net kicks are small.

\begin{figure}
\epsscale{0.7} 
\plotone{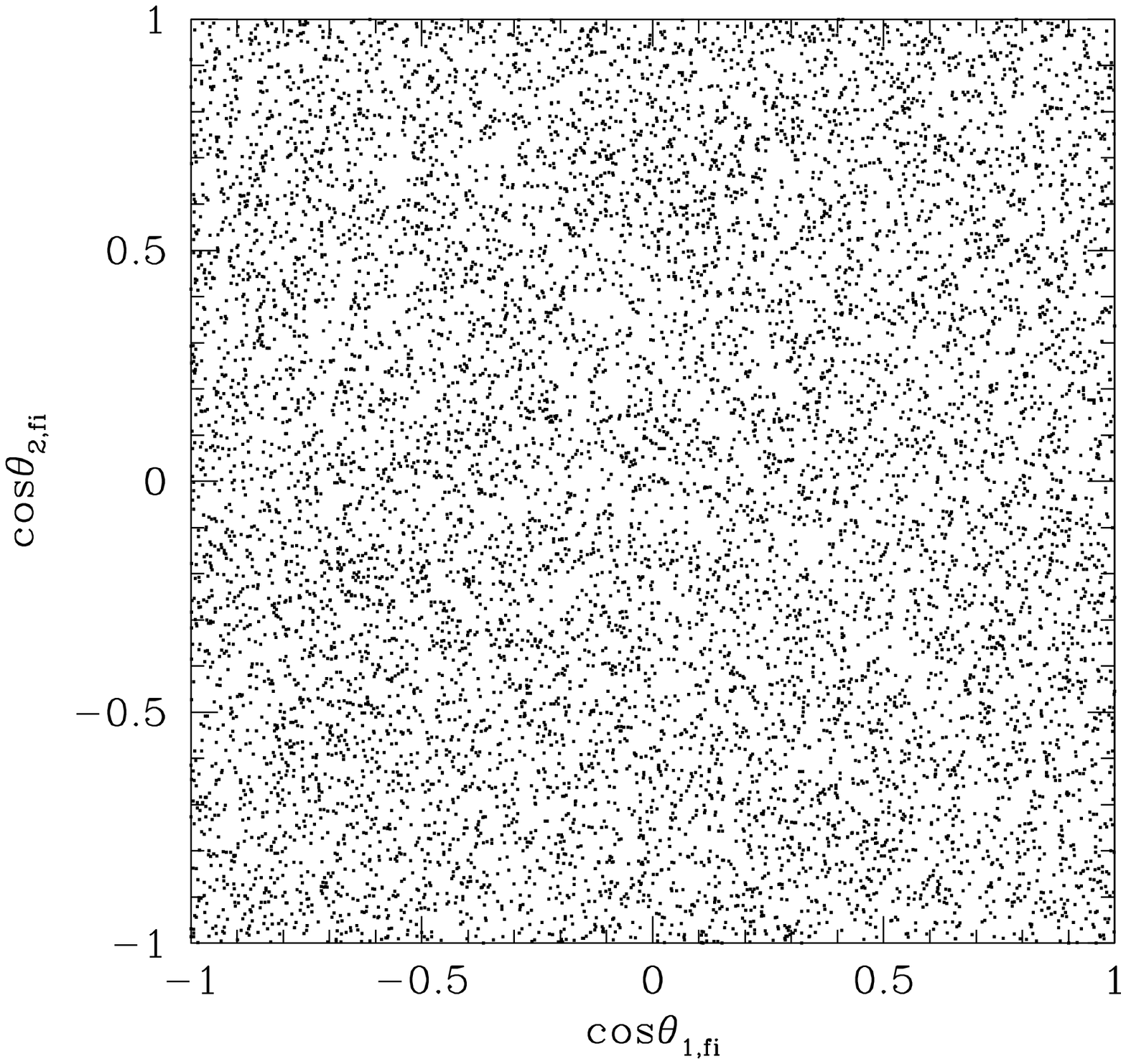}
\caption{Demonstration that in gas-poor mergers there is no
tendency to align black hole spin angles.  Here we show the
distribution of the dot products with the orbital angular momentum
axis of the final spin axis of the larger mass ($\cos\theta_{1,fi}$)
and smaller mass ($\cos\theta_{2,fi}$) black hole, evolved using the
formalism of \cite{Schnittman04}.  We assume that at an initial
separation of $1000m$ (where $m=1$ is the total mass of the binary)
the spin directions and orbital axis are distributed isotropically,
and we integrate inward to $10m$ assuming component masses $m_1=0.55$
and $m_2=0.45$ and dimensionless spin parameters ${\hat a}_1={\hat
a}_2=1$. The final spin angles show no alignment towards each other or
towards the orbital axis, hence other mechanisms are needed to avoid
ejecting merged black holes from their host galaxies.
\label{fig:finalangles}} 
\end{figure}

Using equations A8 and A10 from \cite{Schnittman04}, we have evolved
the angles between the two spin vectors, and between the spins and the
orbital angular momentum. We find that for isotropically distributed
initial spins and orbits, the spins and orbits at close separation are
also close to isotropically distributed (see
Figure~\ref{fig:finalangles}). Thus, although (as we confirm)
\cite{Schnittman04} showed that for special orientations the spins
might align (e.g., for an initial $\cos\theta_1\approx 1$, or as we
also discovered, for an initial $\cos\theta_2\approx -1$), the initial
conditions resulting in such alignment are special and subtend only a
small solid angle.

The conclusion is that gas-poor mergers alone cannot align spins
sufficiently to avoid large kicks due to gravitational radiation
recoil.  Indeed, \cite{sb07} find that for mass ratios $q>0.25$, spin
magnitudes ${\hat a}_1={\hat a_2}=0.9$, and isotropic spin directions,
$\sim 8$\% of coalescences result in kick speeds $>1000$~km~s$^{-1}$
and $\sim 30$\% yield speeds $>500$~km~s$^{-1}$.  The high maximum
speeds inferred by \cite{Campanelli07b} are likely to increase these
numbers. We now discuss gas-rich mergers, which can naturally reduce
the kick speeds by aligning black hole spins with their orbital axis.

\section{Gas-rich Mergers \label{S_wet_mergers}}

Consider now a gas rich environment, which is common in many galactic
mergers.  The key new element is that gas accretion can exert torques
that change the direction but not the magnitude of the spin of a black
hole, and that the lever arm for these torques can be tens of
thousands of gravitational radii \citep{bp75,np98,na99}.  In
particular, \citet{np98} and \cite{na99} demonstrate that the black
hole can align with the larger scale accretion disk on a timescale
that is as short as 1\% of the accretion time. An important ingredient
of this scenario is the realization by \citet{pp83} that the warps are
transmitted through the disk on a timescale that is shorter by a
factor of $1/2\alpha^2$ compared with the transport of the orbital
angular momentum in flat disks, where $\alpha\sim 0.01-0.1$ is the
standard viscosity parameter \citep{ss73}. The question that
distinguishes gas-rich from gas-poor mergers is therefore whether the
accreted mass is $\sim 0.01-0.1M_{\rm bh}$ during the sinking of the
black holes towards the center of the merged galaxy, where $M_{\rm
bh}$ is a black hole mass.

Numerical simulations show that galactic mergers trigger large gas
inflows into the central kiloparsec, which in gas rich galaxies can
result in a $\sim 10^9 \Msun$ central gas remnant with a diameter of
only few$\times$100~pc \citep{bh91,bh96,mh94,msh05,kazantzidis05}. 
Such mergers are thought to be the progenitors of ultraluminous
infrared galaxies. \citet{kazantzidis05} find that the strong gas
inflows observed in cooling and star formation simulations always
produce a rotationally supported nuclear disk of size $\sim 1-2$~kpc
with peak rotational velocities in the range of 250$-$300 ${\rm km\,
s^{-1}}$.

The results of numerical simulations are in good agreement with
observations, which also show that the total mass of the gas
accumulated in the central region of merger galaxies can reach
$10^9-10^{10}\Msun$ and in some cases account for about half of the
enclosed dynamical mass \citep{tacconi99}. Observations imply that the
cold, molecular gas settles into a geometrically thick, rotating
structure with velocity gradients similar to these obtained in
simulations and with densities in the range $10^2-10^5\,{\rm cm^{-3}}$
\citep{ds98}. Both observations and simulations of multiphase
interstellar matter with stellar feedback show a broad range of gas
temperatures, where the largest fraction of gas by mass has a
temperature of about $100\,$K \citep{wn01,wn02}.

We therefore consider an idealized model based on these observations
and simulations. In our model, the two black holes are displaced from
the center embedded within the galactic-scale gas disk. We are mainly
concerned with the phase in which the holes are separated by hundreds
of parsecs, hence the enclosed gas and stellar mass greatly exceeds
the black hole masses and we can assume that the black holes interact
independently with the disk. Based on the results of
\citet{escala04,escala05}, \citet{mayer06}, and \citet*{dotti06} the
time for the black holes to sink from these separations to the center
of the disk due to dynamical friction against gaseous and stellar
background is $\leq5\times10^7$yr, which is comparable to the
starburst timescale, $\sim10^8$yr \citep{larson87}.

The accretion onto the holes is mediated by their nuclear accretion
disks fed from the galactic scale gas disk at the Bondi rate, ${\dot
  M}_{\rm Bondi}$, as long as it does not exceed the Eddington rate,
${\dot M_{\rm Edd}}$ \citep{GT04}. Locally, one can estimate the Bondi
radius $R_{\rm Bondi}=GM_{\rm bh}/v_g^2\approx 40~{\rm pc}\,(M_{\rm
  bh}/10^8\,M_\odot)(v_g/100~{\rm km~s}^{-1})^{-2}$ that would be
appropriate for a total gas speed at infinity, relative to a black
hole, of $v_g$ (we use a relatively large scaling of 100~km~s$^{-1}$
for this quantity to be conservative and to include random motions of
gas clouds as well as the small thermal speed within each cloud).  The
accretion rate onto the holes will then be $\min({\dot M_{\rm
    Bondi}},{\dot M_{\rm Edd}})$, where ${\dot M}_{\rm Bondi}\approx
1~M_\odot\,{\rm yr}^{-1}\; (v_g/100\,{\rm
  km\,s}^{-1})^{-3}(n/100\,{\rm cm}^{-3})(M_{\rm bh}/10^8\Msun)^2$.
We also note that clearing of a gap requires accretion of enough gas
to align the holes with the large-scale gas flow.

At this rate, the holes will acquire 1-10\% of their mass in a time
short compared to the time needed for the holes to spiral in towards
the center or the time for a starburst to deplete the supply of gas.
The gas has significant angular momentum relative to the black holes:
analogous simulations in a planetary formation environment suggest
that the circularization radius is some hundredths of the capture
radius (e.g., \citealt{HB91,HB92}). This corresponds to more than
$10^5$ gravitational radii, hence alignment of the black hole spin
axes is efficient (and not antialignment, since the cumulative angular
momentum of the accretion disk is much greater than the angular
momentum of the black holes; see \citealt{king05,lp06}).

If the black hole spins have not been aligned by the time their Bondi
radii overlap and a hole is produced in the disk, further alignment
seems unlikely \citep[for a different interpretation see][]{liu04}. The
reason is twofold: the shrinking of the binary due to circumbinary
torques is likely to occur within $<{\rm few}\times 10^7$~yr
\citep{escala04,escala05}, and accretion across the gap only occurs at
$\sim 10$\% of the rate it would have for a single black hole
\citep{lsa99,ld06,mm06}, with possibly even smaller rates onto the
holes themselves. This therefore leads to the gas-poor merger
scenario, suggesting that massive ellipticals or ellipticals with slow
rotation or boxy orbits have a several percent chance of having
ejected their merged black holes but that other galaxy types will
retain their holes securely.

\section{Predictions, Discussion and Conclusions}

We propose that when two black hole accrete at least $\sim 1-10$\% of
their masses during a gas-rich galactic merger, their spins will align
with the orbital axis and hence the ultimate gravitational radiation
recoil will be $<200$~km~s$^{-1}$. In this section we discuss several
other observational predictions that follow from this scenario.  The
best diagnostic of black hole spin orientation is obtained by
examining AGN jets.  All viable jet formation mechanisms result
in a jet that is initially launched along the spin axis of the black
hole.  This is the case even if the jet is energized by the accretion
disk rather than the black hole spin since the orientation of the
inner accretion disk will be slaved to the black hole spin axis by the
Bardeen-Petterson effect \citep{bp75}.

At first glance, alignment of black hole spin with the large scale
angular momentum of the gas would seem to run contrary to the
observation that Seyfert galaxies have jets that are randomly oriented
relative to their host galaxy disks \citep{kinney00}.  However,
Seyfert morphology is not consistent with recent major mergers
\citep{veilleux03}, hence randomly-oriented minor mergers or internal
processes (e.g., scattering of a giant molecular cloud into the black
hole loss cone) are likely the cause of the current jet directions in
Seyfert galaxies.

Within our scenario, one will never witness dramatic spin orientation
changes during the final phase of black hole coalescence following a
gas-rich merger. There is a class of radio-loud AGN known as
``X-shaped radio galaxies'', however, that possess morphologies
interpreted precisely as a rapid ($<10^5$\,yr) re-alignment of black
hole spin during a binary black hole coalescence
\citep{ekers78,dt02,wang03,komossa03b,lr07,cheung07}. These sources
have relatively normal ``active'' radio lobes (often displaying jets
and hot-spots) but, in addition, have distinct ``wings'' at a
different position angle.  The spin realignment hypothesis argues that
the wings are old radio lobes associated with jets from the one of the
pre-coalescence black holes in which the spin axis possessed an
entirely different orientation to the post-coalescence remnant black
hole \citep{me02}.  If this hypothesis is confirmed by, for example,
catching one of these systems in the small window of time in which
both sets of radio lobes have active hot spots, it would contradict
our scenario unless it can be demonstrated that all such X-shaped
radio-galaxies originate from gas-poor mergers.

However, the existence of a viable alternative mechanism currently
prevents a compelling case from being made that X-shaped radio sources
are a unique signature of mis-aligned black hole coalescences.  The
collision and subsequent lateral expansion of the radio galaxy
backflows can equally well produce the observed wings
\citep{capetti02}.  Indeed, there is circumstantial evidence
supporting the backflow hypothesis.  \citet{kraft05} present a Chandra
observation of the X-shaped radio galaxy 3C~403 and find that the hot
ISM of the host galaxy is strongly elliptical, with a (projected)
eccentricity of $e\sim 0.6$.  Furthermore, the wings of the ``X'' are
closely aligned with the minor axis of the gas distribution,
supporting a model in which the wings correspond to a colliding
backflow that has ``blown'' out of the ISM along the direction of
least resistance.  Although 3C~403 is the only X-shaped radio galaxy
for which high-resolution X-ray maps of the hot ISM are available,
\citet{capetti02} have noted that a number of X-shaped sources have
wings that are oriented along the minor axis of the {\it optical} host
galaxy.  This suggests that the conclusion of \citet{kraft05} for
3C~403 may be more generally true.

There is one particular system, 0402$+$379, that might provide a
direct view of spin alignment in a binary black hole system.  Very
Long Baseline Array (VLBA) imaging of this radio galaxy by
\citet{maness04} discovered two compact flat spectrum radio cores, and
follow-up VLBA observations presented by \citet{rodriguez06} showed
the cores to be stationary.  A binary supermassive black hole is the
most satisfactory explanation for this source, with the projected
distance between the two black holes being only 7.3\,pc.  Within the
context of our gas-rich merger scenario, these two black holes already
have aligned spins.  Existing VLBA data only show a jet associated
with one of the radio cores.  We predict that, if a jet is eventually
found associated with the other radio core, it will have the same
position angle as the existing jet.

Evidence already exists for alignment of a coalescence remnant with
its galactic scale gas disk. \citet{perlman01} imaged the host
galaxies of three compact symmetric objects and discovered nuclear gas
disks approximately normal to the jet axis. The presence of such a
nuclear gas disk as well as disturbances in the outer isophotes of all
three host galaxies suggests that these galaxies had indeed suffered
major gas-rich mergers within the past $10^8$\,yr.

In conclusion, we propose that in the majority of galactic mergers,
torques from gas accretion align the spins of supermassive black holes
and their orbital axis with large-scale gas disks.  This scenario
helps explain the ubiquity of black holes in galaxies despite the
potentially large kicks from gravitational radiation recoil.  Further
observations, particularly of galaxy mergers that do not involve
significant amounts of gas, will test our predictions and may point to
a class of large galaxies without central black holes.

\acknowledgments

We thank Doug Hamilton and Eve Ostriker for insightful discussions.
TB thanks the UMCP-Astronomy Center for Theory and Computation Prize
Fellowship program for support.  CSR and MCM gratefully acknowledge
support from the National Science Foundation under grants AST0205990
(CSR) and AST0607428 (CSR and MCM).


\clearpage

\end{document}